\documentclass[twocolumn,tighten,trackchanges]{aastex61}
\usepackage{natbib,color,amsmath}
\usepackage[normalem]{ulem}
\usepackage{epstopdf}
\usepackage{graphicx}
\definecolor{red}{rgb}{1.0,0.0,0.0}

\newcommand{\Mj}[1]{$M_\mathrm{Jup}$}

\usepackage[utf8]{inputenc}

\begin{document}
\title{Detection of a low-mass stellar companion to the accelerating A2IV star HR 1645}

\correspondingauthor{Robert J. De Rosa}
\email{rderosa@stanford.edu}

\author[0000-0002-4918-0247]{Robert J. De Rosa}
\affiliation{Kavli Institute for Particle Astrophysics and Cosmology, Stanford University, Stanford, CA 94305, USA}

\author[0000-0001-6975-9056]{Eric L. Nielsen}
\affiliation{Kavli Institute for Particle Astrophysics and Cosmology, Stanford University, Stanford, CA 94305, USA}

\author[0000-0003-0029-0258]{Julien Rameau}
\affiliation{Univ. Grenoble Alpes/CNRS, IPAG, F-38000 Grenoble, France}
\affiliation{Institut de Recherche sur les Exoplan{\`e}tes, D{\'e}partement de Physique, Universit{\'e} de Montr{\'e}al, Montr{\'e}al QC, H3C 3J7, Canada}

\author[0000-0002-5092-6464]{Gaspard Duch\^ene}
\affiliation{Department of Astronomy, University of California, Berkeley, CA 94720, USA}
\affiliation{Univ. Grenoble Alpes/CNRS, IPAG, F-38000 Grenoble, France}

\author[0000-0002-7162-8036]{Alexandra Z. Greenbaum}
\affiliation{Department of Astronomy, University of Michigan, Ann Arbor, MI 48109, USA}

\author[0000-0003-0774-6502]{Jason J. Wang}
\altaffiliation{51 Pegasi b Fellow}
\affiliation{Department of Astronomy, California Institute of Technology, Pasadena, CA 91125, USA}

\author[0000-0001-5172-7902]{S. Mark Ammons}
\affiliation{Lawrence Livermore National Laboratory, Livermore, CA 94551, USA}

\author[0000-0002-5407-2806]{Vanessa P. Bailey}
\affiliation{Jet Propulsion Laboratory, California Institute of Technology, Pasadena, CA 91109, USA}

\author[0000-0002-7129-3002]{Travis Barman}
\affiliation{Lunar and Planetary Laboratory, University of Arizona, Tucson AZ 85721, USA}

\author{Joanna Bulger}
\affiliation{Institute for Astronomy, University of Hawaii, 2680 Woodlawn Drive, Honolulu, HI 96822, USA}
\affiliation{Subaru Telescope, NAOJ, 650 North A{'o}hoku Place, Hilo, HI 96720, USA}

\author[0000-0001-6305-7272]{Jeffrey Chilcote}
\affiliation{Department of Physics, University of Notre Dame, 225 Nieuwland Science Hall, Notre Dame, IN, 46556, USA}

\author[0000-0003-0156-3019]{Tara Cotten}
\affiliation{Department of Physics and Astronomy, University of Georgia, Athens, GA 30602, USA}

\author{Rene Doyon}
\affiliation{Institut de Recherche sur les Exoplan{\`e}tes, D{\'e}partement de Physique, Universit{\'e} de Montr{\'e}al, Montr{\'e}al QC, H3C 3J7, Canada}

\author[0000-0002-0792-3719]{Thomas M. Esposito}
\affiliation{Department of Astronomy, University of California, Berkeley, CA 94720, USA}

\author[0000-0002-0176-8973]{Michael P. Fitzgerald}
\affiliation{Department of Physics \& Astronomy, University of California, Los Angeles, CA 90095, USA}

\author[0000-0002-7821-0695]{Katherine B. Follette}
\affiliation{Physics and Astronomy Department, Amherst College, 21 Merrill Science Drive, Amherst, MA 01002, USA}

\author[0000-0003-3978-9195]{Benjamin L. Gerard}
\affiliation{University of Victoria, 3800 Finnerty Rd, Victoria, BC, V8P 5C2, Canada}
\affiliation{National Research Council of Canada Herzberg, 5071 West Saanich Rd, Victoria, BC, V9E 2E7, Canada}

\author[0000-0002-4144-5116]{Stephen J. Goodsell}
\affiliation{Gemini Observatory, 670 N. A'ohoku Place, Hilo, HI 96720, USA}

\author{James R. Graham}
\affiliation{Department of Astronomy, University of California, Berkeley, CA 94720, USA}

\author[0000-0003-3726-5494]{Pascale Hibon}
\affiliation{Gemini Observatory, Casilla 603, La Serena, Chile}

\author[0000-0001-9994-2142]{Justin Hom}
\affiliation{School of Earth and Space Exploration, Arizona State University, PO Box 871404, Tempe, AZ 85287, USA}

\author[0000-0003-1498-6088]{Li-Wei Hung}
\affiliation{Natural Sounds and Night Skies Division, National Park Service, Fort Collins, CO 80525, USA}

\author{Patrick Ingraham}
\affiliation{Large Synoptic Survey Telescope, 950N Cherry Ave., Tucson, AZ 85719, USA}

\author{Paul Kalas}
\affiliation{Department of Astronomy, University of California, Berkeley, CA 94720, USA}
\affiliation{SETI Institute, Carl Sagan Center, 189 Bernardo Ave.,  Mountain View CA 94043, USA}

\author[0000-0002-9936-6285]{Quinn Konopacky}
\affiliation{Center for Astrophysics and Space Science, University of California San Diego, La Jolla, CA 92093, USA}

\author[0000-0001-7687-3965]{James E. Larkin}
\affiliation{Department of Physics \& Astronomy, University of California, Los Angeles, CA 90095, USA}

\author[0000-0003-1212-7538]{Bruce Macintosh}
\affiliation{Kavli Institute for Particle Astrophysics and Cosmology, Stanford University, Stanford, CA 94305, USA}

\author{J\'er\^ome Maire}
\affiliation{Center for Astrophysics and Space Science, University of California San Diego, La Jolla, CA 92093, USA}

\author[0000-0001-7016-7277]{Franck Marchis}
\affiliation{SETI Institute, Carl Sagan Center, 189 Bernardo Ave.,  Mountain View CA 94043, USA}

\author[0000-0002-5251-2943]{Mark S. Marley}
\affiliation{NASA Ames Research Center, MS 245-3, Mountain View, CA 94035, USA}

\author[0000-0002-4164-4182]{Christian Marois}
\affiliation{National Research Council of Canada Herzberg, 5071 West Saanich Rd, Victoria, BC, V9E 2E7, Canada}
\affiliation{University of Victoria, 3800 Finnerty Rd, Victoria, BC, V8P 5C2, Canada}

\author[0000-0003-3050-8203]{Stanimir Metchev}
\affiliation{Department of Physics and Astronomy, Centre for Planetary Science and Exploration, The University of Western Ontario, London, ON N6A 3K7, Canada}
\affiliation{Department of Physics and Astronomy, Stony Brook University, Stony Brook, NY 11794-3800, USA}

\author[0000-0001-6205-9233]{Maxwell A. Millar-Blanchaer}
\altaffiliation{NASA Hubble Fellow}
\affiliation{Jet Propulsion Laboratory, California Institute of Technology, Pasadena, CA 91109, USA}

\author[0000-0001-7130-7681]{Rebecca Oppenheimer}
\affiliation{Department of Astrophysics, American Museum of Natural History, New York, NY 10024, USA}

\author{David Palmer}
\affiliation{Lawrence Livermore National Laboratory, Livermore, CA 94551, USA}

\author{Jennifer Patience}
\affiliation{School of Earth and Space Exploration, Arizona State University, PO Box 871404, Tempe, AZ 85287, USA}

\author[0000-0002-3191-8151]{Marshall Perrin}
\affiliation{Space Telescope Science Institute, Baltimore, MD 21218, USA}

\author{Lisa Poyneer}
\affiliation{Lawrence Livermore National Laboratory, Livermore, CA 94551, USA}

\author{Laurent Pueyo}
\affiliation{Space Telescope Science Institute, Baltimore, MD 21218, USA}

\author[0000-0002-9246-5467]{Abhijith Rajan}
\affiliation{Space Telescope Science Institute, Baltimore, MD 21218, USA}

\author[0000-0002-9667-2244]{Fredrik T. Rantakyr\"o}
\affiliation{Gemini Observatory, Casilla 603, La Serena, Chile}

\author[0000-0003-1698-9696]{Bin Ren}
\affiliation{Department of Physics and Astronomy, Johns Hopkins University, Baltimore, MD 21218, USA}

\author[0000-0003-2233-4821]{Jean-Baptiste Ruffio}
\affiliation{Kavli Institute for Particle Astrophysics and Cosmology, Stanford University, Stanford, CA 94305, USA}

\author[0000-0002-8711-7206]{Dmitry Savransky}
\affiliation{Sibley School of Mechanical and Aerospace Engineering, Cornell University, Ithaca, NY 14853, USA}

\author{Adam C. Schneider}
\affiliation{School of Earth and Space Exploration, Arizona State University, PO Box 871404, Tempe, AZ 85287, USA}

\author[0000-0003-1251-4124]{Anand Sivaramakrishnan}
\affiliation{Space Telescope Science Institute, Baltimore, MD 21218, USA}

\author[0000-0002-5815-7372]{Inseok Song}
\affiliation{Department of Physics and Astronomy, University of Georgia, Athens, GA 30602, USA}

\author[0000-0003-2753-2819]{Remi Soummer}
\affiliation{Space Telescope Science Institute, Baltimore, MD 21218, USA}

\author{Melisa Tallis}
\affiliation{Kavli Institute for Particle Astrophysics and Cosmology, Stanford University, Stanford, CA 94305, USA}

\author{Sandrine Thomas}
\affiliation{Large Synoptic Survey Telescope, 950N Cherry Ave., Tucson, AZ 85719, USA}

\author[0000-0001-5299-6899]{J. Kent Wallace}
\affiliation{Jet Propulsion Laboratory, California Institute of Technology, Pasadena, CA 91109, USA}

\author[0000-0002-4479-8291]{Kimberly Ward-Duong}
\affiliation{Physics and Astronomy Department, Amherst College, 21 Merrill Science Drive, Amherst, MA 01002, USA}

\author{Sloane Wiktorowicz}
\affiliation{Department of Astronomy, UC Santa Cruz, 1156 High St., Santa Cruz, CA 95064, USA }

\author[0000-0002-9977-8255]{Schuyler Wolff}
\affiliation{Leiden Observatory, Leiden University, 2300 RA Leiden, The Netherlands}


\begin{abstract}
The $\sim500$\, Myr A2IV star HR 1645 has one of the most significant low-amplitude accelerations of nearby early-type stars measured from a comparison of the  {\it Hipparcos} and {\it Gaia} astrometric catalogues. This signal is consistent with either a stellar companion with a moderate mass ratio ($q\sim0.5$) on a short period ($P<1$\,yr), or a substellar companion at a separation wide enough to be resolved with ground-based high contrast imaging instruments; long-period equal mass ratio stellar companions that are also consistent with the measured acceleration are excluded with previous imaging observations. The small but significant amplitude of the acceleration made HR 1645 a promising candidate for targeted searches for brown dwarf and planetary-mass companions around nearby, young stars. In this paper we explore the origin of the astrometric acceleration by modelling the signal induced by a wide-orbit M8 companion discovered with the Gemini Planet Imager, as well as the effects of an inner short-period spectroscopic companion discovered a century ago but not since followed-up. We present the first constraints on the orbit of the inner companion, and demonstrate that it is a plausible cause of the astrometric acceleration. This result demonstrates the importance of vetting of targets with measured astrometric acceleration for short-period stellar companions prior to conducting targeted direct imaging surveys for wide-orbit substellar companions.
\end{abstract}

\section{Introduction}
Plane-of-sky measurements of a star's position relative to distant background stars can be used to monitor the reflex motion of the target in response to an unseen orbiting companion. Several companion searches following this astrometric technique have been carried out from the ground \citep[e.g.,][]{Sahlmann:2010hh} and from space \citep[e.g.,][]{Benedict:1999fn}, leading to the detection of several stellar and substellar companions \citep[e.g.,][]{Pravdo:2005kt,Goldin:2007bl,Reffert:2011ca,Sahlmann:2013ks}. Because the photocenter displacement increases with orbital period, precise absolute astrometry over a long time baseline has the potential to reveal populations of stellar, substellar, and planetary-mass companions to nearby stars that are inaccessible to current high-contrast imaging instruments. This was achieved with the first space-based astrometric mission \textit{Hipparcos} \citep{Perryman:1997vj}, and to a lesser extent with the \textit{Hubble Space Telescope}. When combined with radial velocity observations, these measurements allowed for a determination of the full three-dimensional orbit, and for a direct measurement of the mass of the orbiting companion \citep[e.g.,][]{Benedict:2010iu,Sahlmann:2010hh}.

Looking ahead, the \textit{Gaia} mission will have the precision necessary to reveal thousands of exoplanets over its lifetime \citep{Casertano:2008es,Perryman:2014jr}. While we await the release of the final {\it Gaia} catalogue, the twenty-four years that separate \textit{Hipparcos} from \textit{Gaia} provide a baseline that is long enough to detect the acceleration of the proper motion of a star due to an substellar companion on an orbit that is wide enough to be directly imaged with current ground-based high contrast imaging instruments (e.g., \citealp{Kervella:2019bw}). Indeed, the acceleration inferred from the two catalogues has already been successfully combined with long-term radial velocity measurements to obtain precise dynamical mass measurements of several substellar companions \citep{Snellen:2018gc,Brandt:2018uj,Dupuy:2019cy}.

In this paper we report on the discovery of a wide (32\,au projected separation) late M-type companion to HR 1645 resolved with high-contrast imaging observations obtained with the Gemini Planet Imager (GPI; \citealp{Macintosh:2014js}). This star exhibits a significant acceleration over the 24.25-year baseline between the {\it Hipparcos} and {\it Gaia} missions. While the magnitude of the astrometric acceleration is consistent with the mass of the companion inferred from evolutionary models, the direction is not. Instead, a plausible cause of the astrometric acceleration is a short-period spectroscopic companion discovered in the 1920s but without subsequent follow-up observations (see Section~\ref{sec:rv}). We place the first constraints on the spectroscopic orbit and investigate how this short-period binary could significantly bias the proper motion measurements for this star in both the {\it Hipparcos} and {\it Gaia} catalogues.

\section{HR 1645 -- An Accelerating Early-type Star}
HR 1645 (HIP 23554, Gaia DR2 2960561059245715968) is an A2IV \citep{Houk:1988wv} star at a distance of $59.6\pm0.3$\,pc \citep{GaiaCollaboration:2018io}. The age of the star has previously been estimated through a comparison to evolutionary models as $434\pm34$\,Myr \citep{Zorec:2012il} and  $462_{-65}^{+109}$\,Myr \citep{David:2015ia}. We find a slightly older age of $530_{-140}^{+135}$\,Myr using evolutionary models that account for the rapid rotation of early-type stars \citep{Nielsen:2019cb}, consistent with these estimates. The star is not thought to be a member of any nearby kinematic association. The star does not exhibit a significant infrared excess based on 12 and 24\,{\micron} photometry from the {\it WISE} catalogue \citep{Cutri:2014wx}. Although searches for companions to HR 1645 have ruled out the presence of stellar companions exterior to $\sim1$\,arcsec \citep{DeRosa:2014db}, the discrepancy between the proper motion of HR 1645 reported in the {\it Hipparcos} and {\it Gaia} catalogues provides strong evidence for a massive orbiting companion interior to the detection limits of previous searches.

\subsection{Absolute astrometry}
\label{sec:accl}
\begin{figure}
\includegraphics[width=1.0\columnwidth]{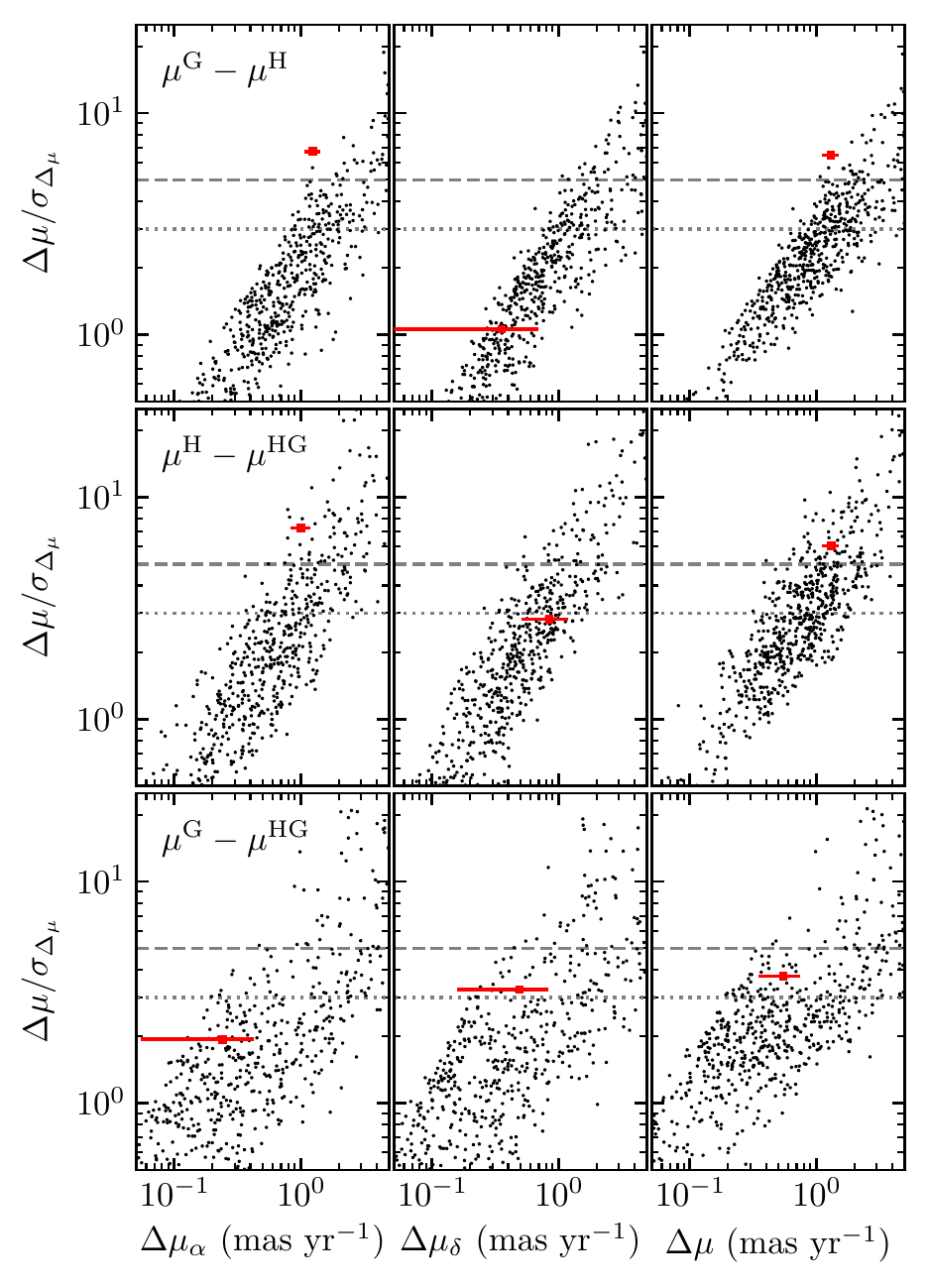}
\caption{Proper motion accelerations, and corresponding significance, measured for a sample of $\sim$700 nearby early-type stars from absolute astrometry within the {\it Hipparcos} and {\it Gaia} catalogues. The columns show the proper motion differential in the right ascension (left column) and declination (middle) directions, as well as the total proper motion difference (right). The rows shows the differential measured from a comparison of the {\it Hipparcos} and {\it Gaia} proper motions (top row), and from the absolute position of the star in the two catalogues and the {\it Hipparcos} (middle) and {\it Gaia} (bottom) proper motions. HR 1645 is indicated (red square), as well as the $3\sigma$ (dashed) and $5\sigma$ (dotted) limits.\label{fig:pm_diff}}
\end{figure}
\label{sec:astrometry}
\begin{deluxetable}{ccccc}
\tablecaption{Astrometric measurements of HR 1645\label{tbl:astrometry}}
\tablehead{\colhead{Property} & \colhead{Unit} & \colhead{Value} & \colhead{Uncertainty} & \colhead{Ref.}}
\startdata
\multicolumn{5}{c}{{\it Hipparcos}}\\
\hline
$\alpha$ & deg & $75.97189716$ & $\pm0.11$\,mas\tablenotemark{a} & 1\\
$\delta$ & deg & $-24.38805710$ & $\pm0.23$\,mas & 1\\
$\mu_{\alpha^{\star}}$ & mas\,yr$^{-1}$ & $26.30$ & $\pm0.14$ & 1\\
$\mu_{\delta}$ & mas\,yr$^{-1}$ & $-38.52$ & $\pm0.30$ & 1\\
$\pi$ & mas & $17.19$ & $\pm0.31$ & 1\\
\hline
\multicolumn{5}{c}{{\it Gaia} DR2}\\
\hline
$\alpha$ & deg & $75.97208419177$ & $\pm0.0508$\,mas\tablenotemark{a} & 2\\
\nodata & \nodata & \nodata & $\pm0.0571$\,mas\tablenotemark{a, b} & 3\\
$\delta$ & deg & $-24.38831085585$ & $\pm0.0597$\,mas & 2\\
\nodata & \nodata & \nodata & $\pm0.0664$\,mas\tablenotemark{b} & 3\\
$\mu_{\alpha^{\star}}$ & mas\,yr$^{-1}$ & $24.958$ & $\pm0.109$ & 2\\
\nodata & \nodata & $25.046$\tablenotemark{b} & $\pm0.124$\tablenotemark{b} & 3\\
$\mu_{\delta}$ & mas\,yr$^{-1}$ & $-38.219$ & $\pm0.135$ & 2\\
\nodata & \nodata & $-38.163$\tablenotemark{b} & $\pm0.151$\tablenotemark{b} & 3\\
$\pi$ & mas & $16.869$ & $\pm0.083$ & 2\\
\nodata & \nodata & \nodata & $\pm0.092$\tablenotemark{b} & 3\\
\enddata
\tablerefs{(1) \citealt{vanLeeuwen:2007dc}; (2) \citealt{GaiaCollaboration:2018io}; (3) this work}
\tablenotetext{a}{Uncertainty in $\alpha^{\star} = \alpha \cos\delta$}
\tablenotetext{b}{After correcting for {\it Gaia} bright star reference frame rotation and the internal to external error ratio}
\end{deluxetable}
The differences in the proper motions of stars between the {\it Hipparcos} and {\it Gaia} epochs, and the proper motion inferred from their positions within each catalogue, are potentially a powerful tool for identifying targets for direct imaging surveys to search for wide orbit substellar companions to nearby, young stars (e.g., \citealp{Brandt:2018dj,Kervella:2019bw}). HR 1645 has one of the most significant ($>3\sigma$) low-amplitude ($\lesssim1$\,mas\,yr$^{-1}$) proper motion differences between the two catalogues of the ~700 A and B type stars within 75 pc (Fig.~\ref{fig:pm_diff}), making it a promising target for such searches. Significant deviations with larger amplitudes are also found for many stars, but these are indicative of more massive stellar or degenerate companions. The proper motion of HR~1645 was measured by {\it Hipparcos} to be $\mu_{\rm H} = (26.30\pm0.14, -38.52\pm0.30)$\,mas\,yr$^{-1}$, and by {\it Gaia} (after correction for the rotation of the bright star reference frame, \citealp{Lindegren:2018gy,Kervella:2019bw}) to be $\mu_{\rm G} = (25.046\pm0.124, -38.163\pm0.151)$\,mas\,yr$^{-1}$, where the proper motions are expressed in the $\alpha^\star=\alpha\cos\delta$ and $\delta$ directions. A significant (6.7$\sigma$) acceleration is measured in the $\alpha^{\star}$ direction, with $\mu_{\rm G}-\mu_{\rm H}=(-1.26\pm0.19, 0.36\pm0.34)$\,mas\,yr$^{-1}$.

The instantaneous position of the star at the reference epoch for both missions was also used to calculate a proper motion over the 24.25-year baseline between the two missions of $\mu_{HG} = (25.2882\pm0.0052, -37.6704\pm0.0098$\,mas\,yr$^{-1}$). There were significant differences between this long-term proper motion and the measurements from {\it Hipparcos} and {\it Gaia} missions. We calculated $\mu_{\rm H}-\mu_{\rm HG}=(1.01\pm0.14, -0.85\pm0.30)$\,mas\,yr$^{-1}$, a $7.2\sigma$ difference in the $\alpha^{\star}$ direction, and $\mu_{\rm G}-\mu_{\rm HG}=(-0.24\pm0.12, -0.49\pm0.15)$\,mas\,yr$^{-1}$, a $3.3\sigma$ difference in the $\delta$ direction. The astrometric measurements from both catalouges are given in Table~\ref{tbl:astrometry}.

\subsection{Inferred companion properties}
\label{sec:accl_fit}
We have developed a framework to predict the masses of companions responsible for measured astrometric accelerations of nearby stars. This astrometric model consisted of eleven free parameters. Seven define the astrometric orbit; the total semi-major axis $a$ ($ = a_1 + a_2$), inclination $i$, eccentricity $e$, argument of periastron $\omega$, longitude of the ascending node $\Omega$, epoch of periastron $\tau$ (in fractions of the orbital period), and the mass of the companion $M_2$ ($M_1$ is held constant at $1.9$\,M$_{\odot}$, based on our fit to evolutionary models with the SED of the star). Two defined the system proper motion in right ascension ($\mu_{\alpha^{\star}}$) and declination ($\mu_{\delta}$), and two accounted for the uncertainty in the position of the photocenter at the {\it Hipparcos} reference epoch of 1991.25  ($\Delta\alpha_0^{\star}$, $\Delta\delta_0$). We fixed the parallax to the {\it Gaia} value of 16.869\,mas, and the radial velocity to the systemic velocity measured in Section~\ref{sec:rv}. In this model we defined the $V$-band photocenter of the system at the {\it Hipparcos} epoch (1991.25) to be ($\alpha_0$, $\delta_0$). We used $V$ as a proxy for the {\it Hipparcos} photometric band $H_{\rm p}$ as the $V-H_{\rm p}$ color for early-type main sequence stars is near zero. The offset between the photocenter measured by {\it Hipparcos} and the system barycenter at this epoch ($\alpha_{0, {\rm b}}$, $\delta_{0, {\rm b}}$) was computed from the remaining free parameters in the model as
\begin{equation}
\begin{split}
\alpha_{0, {\rm b}} &= \alpha_0 + \left[ \Delta\alpha^{\star}_0 + f_{\alpha}\left(a, e, i, \omega, \Omega, \tau, M_2\right)\right]/\cos\delta_{0, b}\\
\delta_{0, {\rm b}} &= \delta_0 + \Delta\delta_0 + f_{\delta}\left(a, e, i, \omega, \Omega, \tau, M_2\right)
\end{split}
\end{equation}
where $f(\ldots)$ is a function that calculates the offset between the barycenter and photocenter of the system in the $\alpha$ and $\delta$ directions from the Keplerian elements \citep{Green:1985ve}. The semi-major axis of the orbit of the primary around the barycenter was defined as $a_1 = Ba$, where $B=M_2/(M_1+M_2)$, and the semi-major axis of the photocenter orbit around the barycenter was defined as $a_p = (B-\beta)a$, where $\beta = (1 + 10^{0.4\Delta m})^{-1}$, and $\Delta m$ was the magnitude difference between the primary and secondary. When the mass ratio $M_2/M_1=q$ is small, the contrast is large and $\beta$ becomes negligible so that $a_p\approx a_1$. Flux ratios were calculated using empirical mass-magnitude relationships \citep{Pecaut:2013ej} for stellar companions, and $\beta$ was assumed to be zero for substellar companions. At 500 Myr, an 80 M$_{\rm Jup}$ brown dwarf is $\sim$16 mags fainter than an A2 star in the $V$ band \citep{Allard:2001fh,Chabrier:2000hq}.

The position of the barycenter was propagated to the {\it Gaia} epoch (2015.5; $\alpha_{1, {\rm b}}$, $\delta_{1, {\rm b}}$) using the formalism described in \citet{Butkevich:2014jt} to account for for the non-rectilinear nature of the equatorial coordinate system and perspective effects over the 24.25-year baseline between the {\it Hipparcos} and {\it Gaia} missions. The offset between the barycenter and the {\it Gaia} $G$-band photocenter ($\alpha_1$, $\delta_1$) was calculated as previously. The instantaneous proper motion of the photocenter was calculated at the {\it Hipparcos} ($\mu_{\alpha^{\star}, 0}$, $\mu_{\delta, 0}$) and {\it Gaia} ($\mu_{\alpha^{\star}, 1}$, $\mu_{\delta, 1}$) epochs which we assume to be equal to the average proper motion over the full \textit{Hipparcos} and \textit{Gaia} DR2 baselines.

We used the parallel-tempered affine-invariant Markov chain Monte Carlo ensemble sampler {\tt emcee} \citep{ForemanMackey:2013io} to sample the posterior distributions of the eleven free parameters in this model. At each step a likelihood was computed as $\ln {\mathcal L} = -\chi^2/2$, where
\begin{equation}
    \chi^2 = R_{\rm H}^\top {\mathbf C}_{\rm H}^{-1}R_{\rm H} + R_{\rm G}^\top {\mathbf C}_{\rm G}^{-1}R_{\rm G}
\end{equation}
with H and G subscripts denoting astrometric measurements from the {\it Hipparcos} and {\it Gaia} catalogues, respectively, and the residual vectors
\begin{equation}
\begin{split}
    R_{\rm H} = &[\Delta\alpha^{\star}_0, \Delta\delta_0,\\
    &\mu_{\alpha^{\star}} + \mu_{\alpha^{\star},0} - \mu_{\alpha^{\star},{\rm H}}, \mu_\delta + \mu_{\delta,0} - \mu_{\delta,{\rm H}}]\\
    R_{\rm G} = &[(\alpha_1 - \alpha_{\rm G})\cos\delta_1, \delta_1 - \delta_{\rm G},\\
    &\mu_{\alpha^{\star}} + \mu_{\alpha^{\star},1} - \mu_{\alpha^{\star},{\rm G}}, \mu_\delta + \mu_{\delta,1} - \mu_{\delta,{\rm G}}] 
\end{split}
\end{equation}
and the covariance matrices ${\mathbf C}_{\rm H}$ and ${\mathbf C}_{\rm G}$ for the {\it Hipparcos} and {\it Gaia} measurements. ${\mathbf C}_{\rm H}$ was computed from the weight matrix ${\mathbf U}$ obtained from the {\it Hipparcos} catalogue using the procedure described in \citet{Michalik:2014eu}, while ${\mathbf C}_{\rm G}$ was computed directly from the correlation coefficients given in the {\it Gaia} catalogue. The rows and columns corresponding to the parallax covariance were removed as this parameter was not a free parameter in this model.

Standard priors on the orbital elements were assumed; uniform in $\log a$, $\cos i$, $e$, $\omega+\Omega$, $\omega-\Omega$, $\tau$. We used a uniform prior for the companion mass between 0--1.0\,$M_{\odot}$, the system proper motion between $-250$ and $250$\,mas\,yr$^{-1}$, and the two offset terms ($\Delta\alpha_0^{\star}$, $\Delta\delta_0$) between $-100$ and $100$\,mas\,yr$^{-1}$. 512 chains were initialized randomly throughout parameter space at 16 different temperatures. The chains were advanced for $10^6$ steps, with the first half being discarded as a ``burn-in''. The chains appeared to be converged based on a visual inspection of the chains, their auto-correlation, and the evolution of the median and 1-$\sigma$ credible interval for each parameter.

\begin{figure}
\includegraphics[width=1.0\columnwidth]{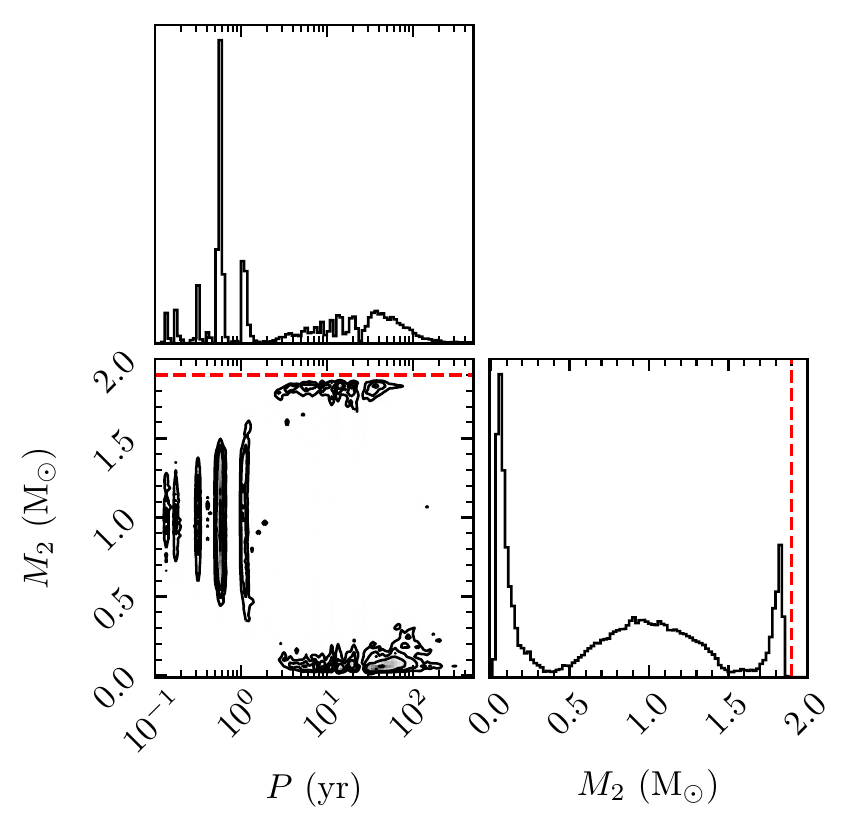}
\caption{Posterior distributions (diagonal) and covariance (lower corner) for the period and mass of the companion inferred from the astrometric accelerations given in \S~\ref{sec:accl}. The red dashed line denotes the assumed mass of the primary (1.9\,$M_{\odot}$). \label{fig:corner}}
\end{figure}
\begin{figure}
\includegraphics[width=1.0\columnwidth]{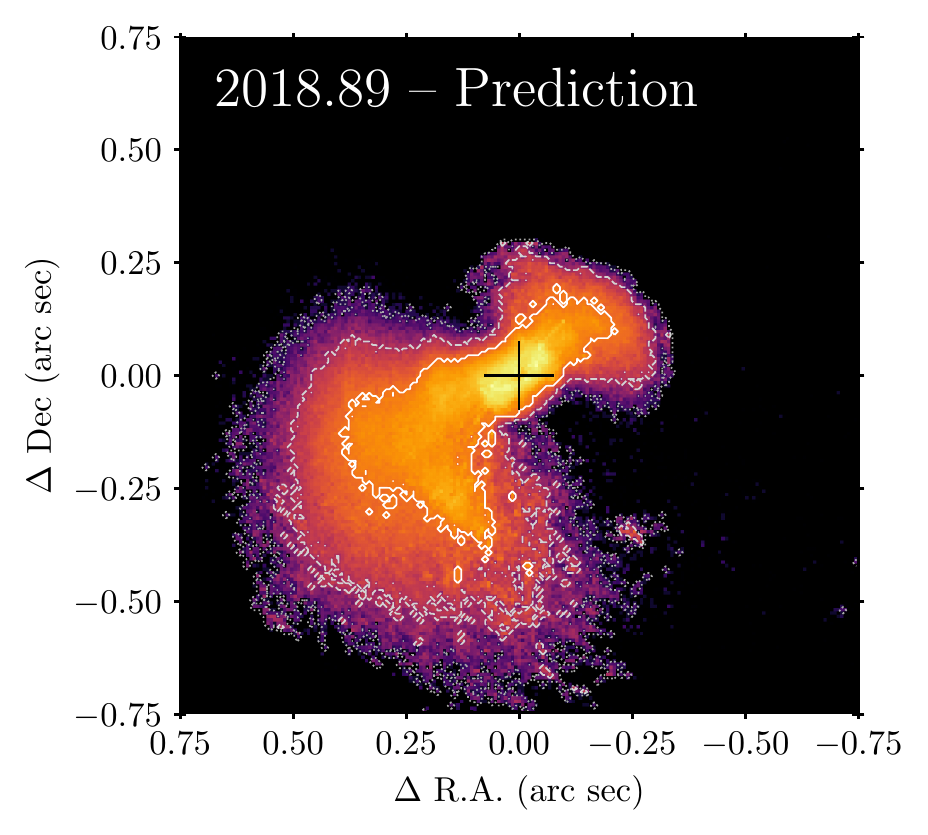}
\caption{Two-dimensional histogram of the predicted location of of the companion at 2018.89 from the MCMC fit to the astrometric signal. The color scale is logarithmic to highlight regions of low probability. Countours denote $1\sigma$ (white solid), $2\sigma$ (white dashed), and $3\sigma$ (gray dotted) credible regions. \label{fig:prediction}}
\end{figure}

Posterior distributions for the period and mass of the companion are shown in Figure~\ref{fig:corner}. The astrometric acceleration is consistent with companions in three distinct regions of mass-period space; near equal-mass ($q\sim1$) companions with long periods ($P>1$\,yr), more intermediate mass ratio ($q\sim0.5$) companions on shorter periods ($P<1$\,yr), and more extreme mass ratio ($q<0.1$) companions on long orbital periods, with masses extending well into the substellar regime. Long-period high-mass companions are excluded with previous high-contrast imaging observations of this star \citep{DeRosa:2014db}. The two remaining possibilities, a short-period intermediate mass ratio stellar companion or a longer-period lower-mass companions, were investigated with dedicated high-contrast imaging observations and an analysis of literature radial velocities for the host star.

\section{HR 1645 B}
\subsection{High-contrast imaging observations}
\label{sec:data}
\begin{figure}
\includegraphics[width=1.0\columnwidth]{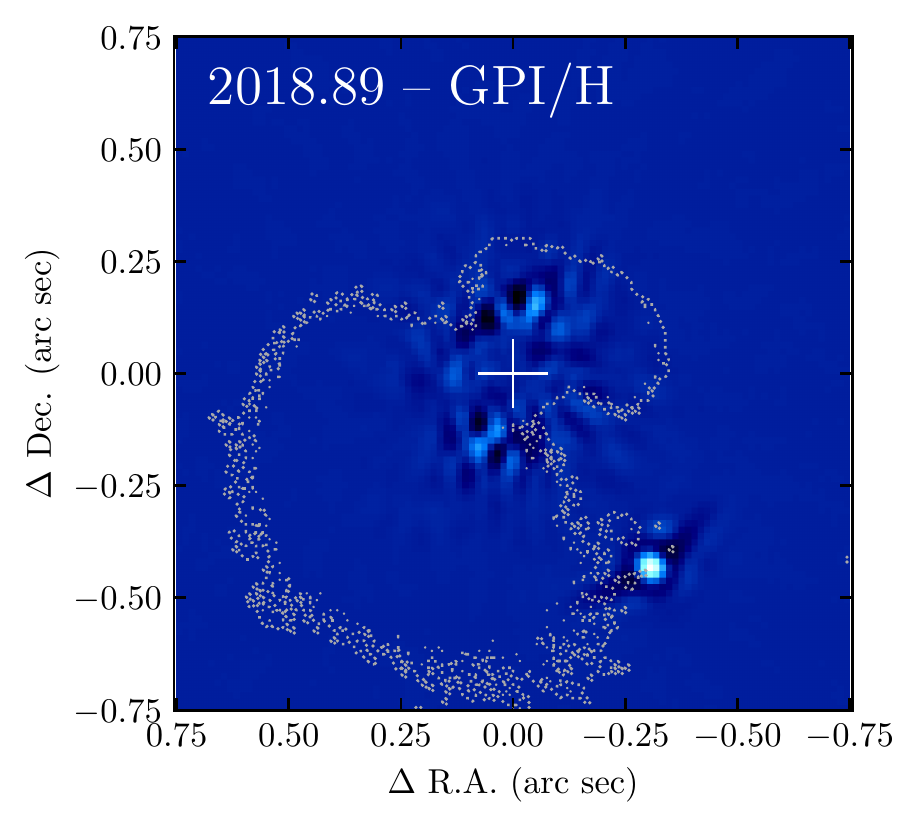}
\caption{cADI reduction of the 2018 November 21 GPI dataset of HR 1645. The companion is clearly detected to the south west of the host star. The $3\sigma$ credible region from Figure~\ref{fig:prediction} is overplotted. \label{fig:imageGPI}}
\end{figure}

HR 1645 was observed as a part of the Gemini Planet Imager Exoplanet Survey\footnote{Gemini program code GS-2017B-Q-500} (GPIES; \citealp{Nielsen:2019cb}) with the Gemini Planet Imager (GPI; \citealp{Macintosh:2014js}) on 2018 November 21 under good conditions. Follow-up observations were carried out on 2019 February 15. For each dataset, the raw data were processed through our automated data reduction pipeline \citep{Wang:2018co}, which uses the GPI Data Reduction Pipeline (DRP; \citealp{Perrin:2014jh}) to perform basic image reduction. Briefly, the DRP subtracts dark background, interpolates bad pixels, converts the 2-D frame into a 3-D ($x, y, \lambda$) cube, interpolates the cube onto a common wavelength axis, corrects for spatial distortion, and identifies the location of the four satellite spots---fiducial replicas of the central star---to measure the position of the star behind the coronagraph.

The 3-D datacubes were further processed to remove the residual PSF of the central star not suppressed by the coronagraph. Large-scale and slowly-varying structures were removed using an apodized Fourier high-pass filter with a cutoff frequency of four units per cycle. The four satellite spots were extracted from each frame and averaged together and over the sequence to build up a template point spread function (PSF) for each wavelength slice. An Angular Differential Imaging-based algorithm (cADI; \citealp{Marois:2006df}) was applied to subtract the residual stellar halo. All frames were rotated to align north with the vertical axis and combined with a trimmed mean ($10\%$) in the temporal direction, resulting in 37 images used to extract the spectrum, and then combined in the spectral dimension to produce a single broad-band image (see Figure \ref{fig:imageGPI}).

The broad-band image was used to measure the position and broad-band contrast of HR 1645 B\footnote{We use the designation B for this wide companion and AaAb for the inner spectroscopic binary discussed later in the manuscript} using the negative forward-model technique \citep{Marois:2010hs, Lagrange:2010fs}. The template PSF was injected in the raw datacubes at a trial position but opposite flux of HR 1645 B and the cADI algorithm was repeated to obtain the residual broad-band image. The process was iterated over these three parameters to minimize the square of the integrated pixel flux in a wedge of $3\times3$ full width at half maximum centered at the trial position. Best fit position and broad-band contrast were obtained with the amoeba-simplex optimization algorithm \citep{Nelder:1965in}. Measurement uncertainties were estimated from independent injections of the template PSF at the best fit separation and contrast but twenty uniformly distributed---besides the spiders---position angles. The fitting process was repeated for each simulated source.

Errors were calculated from the statistical dispersion of the three parameters over the twenty injections. On-chip astrometric measurements were converted into on-sky measurements using the plate scale ($14.161\pm0.021$\,mas\,px$^{-1}$) and north angle correction ($\theta_{\rm true}-\theta_{\rm measured} = 0\fdg45\pm0\fdg11$; \citealp{2019arXiv191008659D}). Errors on the companion astrometry ($0.05$ pixel and $0\fdg09$ for the 2018-11-21 dataset and $0.05$ pixel and $0\fdg1$ for the 2019-02-15 dataset) and star registration ($0.7$ mas; \citealp{Wang:2014ki}) were combined in quadrature. The star-to-satellite spot ratio of $9.39\pm0.01$ mag \citep{Maire:2014gs} was used to calibrate the broad-band contrast and propagate the uncertainty likewise. We measure a separation of $532.9\pm1.3$\,mas, a position angle of $216\fdg99\pm0\fdg13$, and a broadband contrast of $\Delta H=8.13\pm0.01$ mag between HR 1645 B and A in the 2018-11-21 dataset, and $532.8\pm1.3$\,mas, $217.17\pm0\fdg13$, and $8.14\pm0.01$\,mag for the 2019-02-15 dataset. Although we do not detect significant curvature of the orbit of the companion over this short baseline, the measured separation and position angle on the second epoch are $4.5\sigma$ and $5.2\sigma$ discrepant, respectively, from the predicted position of a stationary background object. We used the Besan\c{c}on galactic population model \citep{Robin:2003jk} to estimate the probability of finding a physically unassociated star of the same apparent $H$-band magnitude or brighter within $0\farcs5331$ of HR 1645 to be approximately $3.6\times10^{-5}$. 

The contrast---and associated error--per wavelength slice was measured following the procedure described previously. The flux was the only parameter allowed to vary since the astrometry of the injected template---one per slice---was fixed at the best fit position of HR 1645 B from the previous analysis performed on the wavelength-averaged image. The spectrum of HR 1645 B was obtained by multiplying the contrast with the star-to-satellite spot ratio described previously and the spectrum of the central star derived from a joint fit of synthetic stellar spectra \citep{Castelli:2004ti} and evolutionary models \citep{Paxton:2010jf} to {\it Gaia} and 2MASS photometry.

\subsection{Companion properties}
\label{sec:comparison}
\subsubsection{Spectral type}
\begin{figure*}
\includegraphics[width=1.0\columnwidth,clip, trim=3.3cm 3cm 5.2cm 11cm]{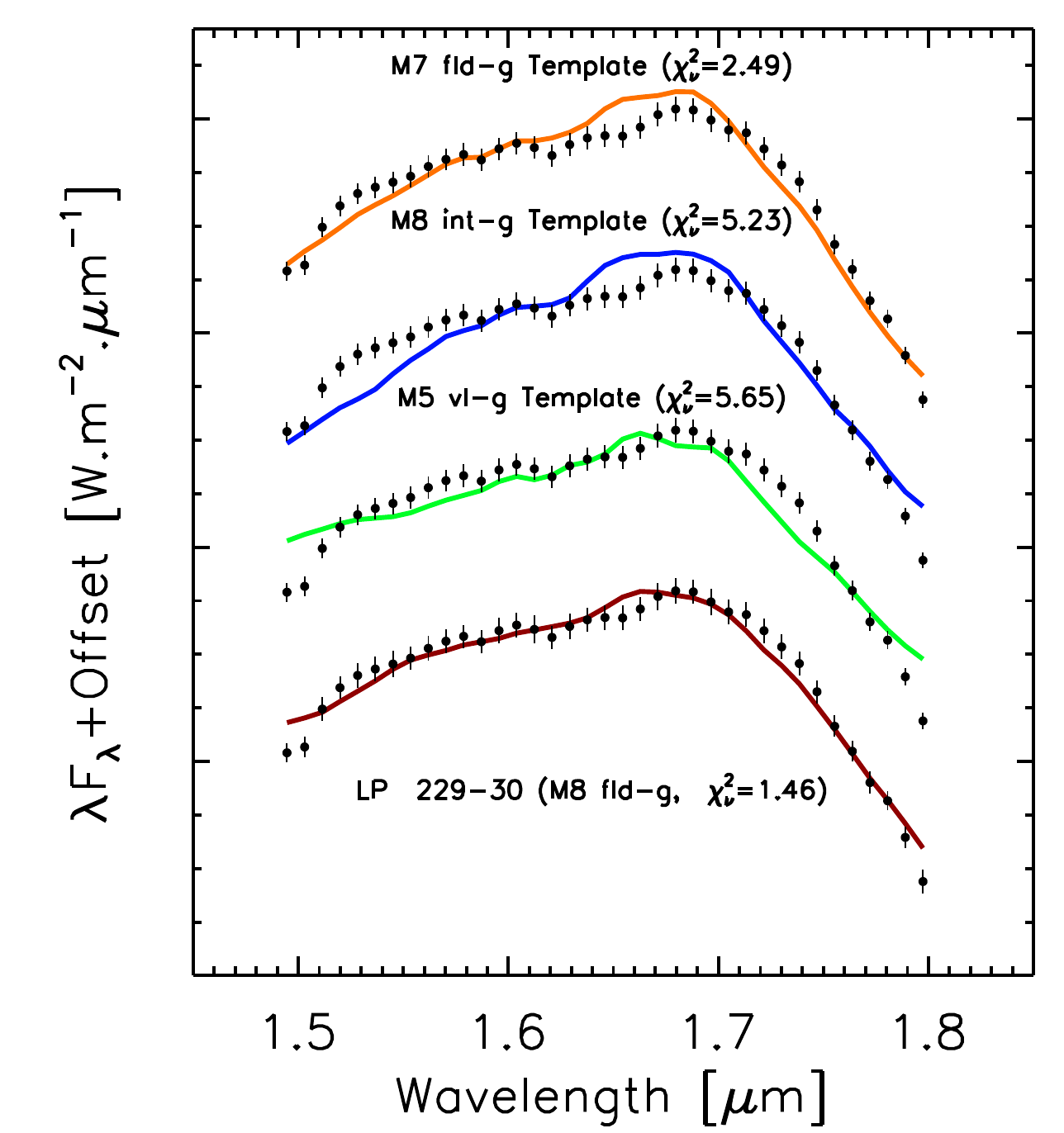}\includegraphics[width=1.0\columnwidth,clip, trim=3.3cm 3cm 5.2cm 11cm]{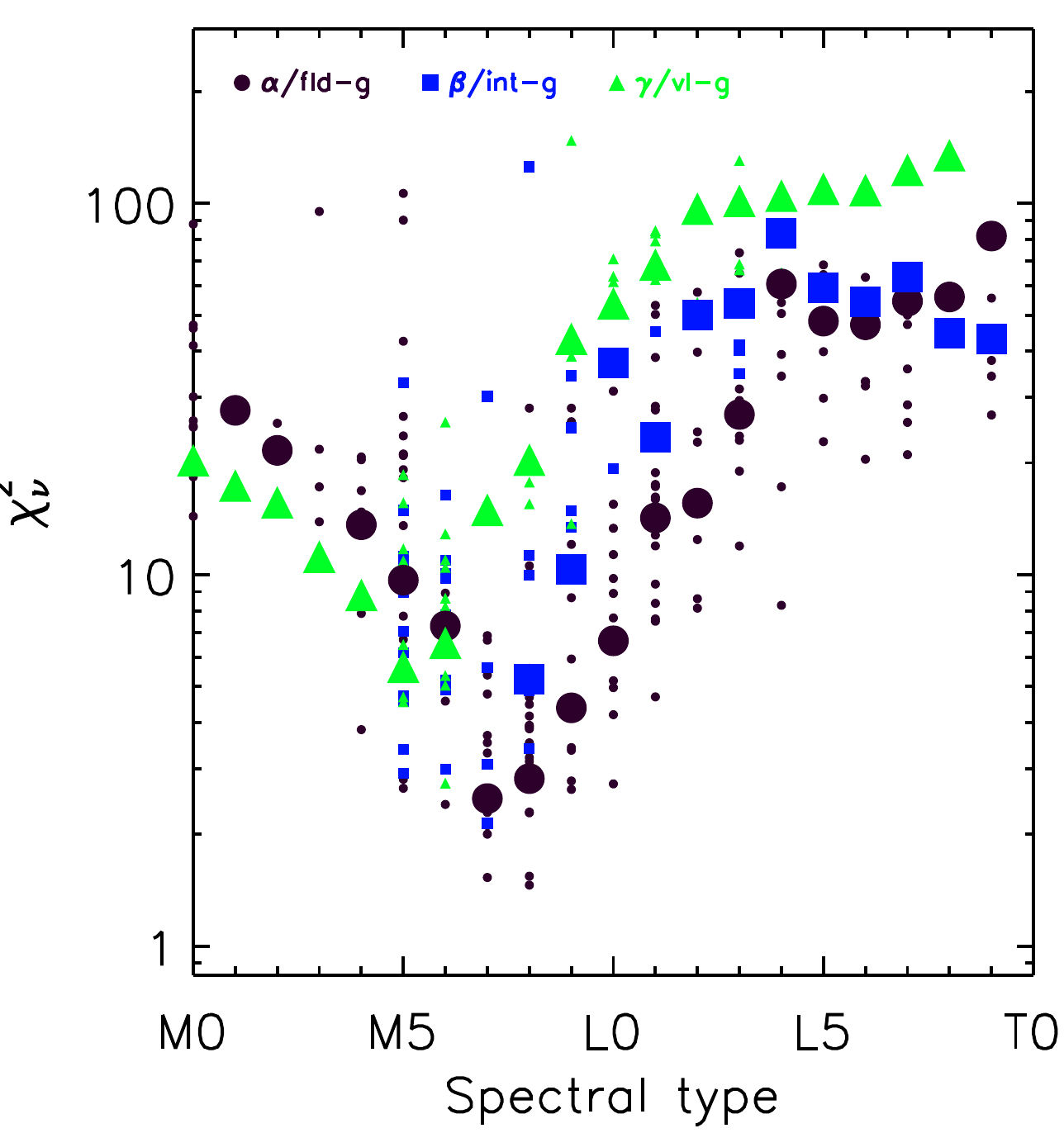}
\caption{\textbf{Left:} The spectrum of HR 1645 B (black points) compared to the average templates for each gravity class (top three rows) and to the best fit individual object from the library. \textbf{Right:} $\chi^2_\nu$ as a function of spectral type for the comparison of the H-band spectrum with a  library of stars and brown-dwarfs. Symbols denote gravity class, with larger symbols denoting the average templates derived for each spectral type and gravity class. \label{fig:spectra}}
\end{figure*}

We compared the $H$ band spectrum of HR 1645 B to a library of near-IR spectra. The library is a compilation of 1164 low- ($R\simeq75$) and medium- ($R\simeq200$) resolution spectra of stars and brown-dwarfs from the Brown Dwarfs in New York City database\footnote{\url{http://database.bdnyc.org/query}} \citep{Filippazzo:2016go,Rodriguez:2016hz}, the IRTF Spectral Library\footnote{\url{http://irtfweb.ifa.hawaii.edu/~spex/IRTF_Spectral_Library/}} \citep{Cushing:2005ed}, the Montr\'eal Spectral Library\footnote{\url{https://jgagneastro.com/the-montreal-spectral-library/}} \citep{Gagne:2015dc,Robert:2016gh}, the SpeX Prism library\footnote{\url{http://pono.ucsd.edu/~adam/browndwarfs/spexprism/}} \citep{Burgasser:2014tr}, and from \citet{Mace:2013jh}, \citet{Best:2015em, Best:2017ih}, and \citet{Leggett:1996eb,Leggett:2017ee}. The library spans spectral types from M to Y at field, intermediate, and very low surface gravity. Spectra at lower signal-to-noise ratio than that of HR 1645 B were discarded from the library. Spectral templates, built from the average of several objects within a given spectral type and gravity class, were added to the library when available from \citet{Luhman:2017jo}, \citet{Gagne:2015dc}, and \citet{Cruz:2018ih}. All spectra were convolved with a Gaussian to degrade their resolution to that of GPI at $H$-band ($R\simeq45$) and interpolated over the same wavelength grid. To compute the $\chi^2$ for each comparison spectrum, its associated errors were added in quadrature to that of HR 1645 B and the minimization factor was calculated analytically. Figure \ref{fig:spectra} shows $\chi^2_\nu$ for M-to-L-type objects, sorted according to the three gravity classes. The $\chi^2_\nu$ distribution is minimal in the M7-M8 range, with best fit from LP 229-30 (M8 fl-g, $\chi^2_\nu=1.46$, \citealt{Cruz:2018ih}), 2MASSI J00034227-28224100 (M7 fl-g, $\chi^2_\nu=1.53$, \citealt{Cruz:2018ih}) and 2MASS J10454932+1254541 (M8 fl-g, $\chi^2_\nu=1.54$, \citealt{Kirkpatrick:2010dc}). Field-gravity objects provide a better fit to the H-band spectrum of HR 1645 B, with a minimum $\chi^2_\nu$ of 1.46, compared to 2.14 at intermediate gravity (2MASS J03350208+2342356, M8 int-g, \citealt{Gagne:2015dc}) and 2.74 at very-low gravity (2MASS J104552630-28193032, M6 vl-g, \citealt{Gagne:2015dc}). This is consistent with the shape of the spectrum being more rounded than typical triangular spectra of young objects. The same trends are observed for template spectra, with field M7-8 being favored ($\chi^2_\nu=2.49$) over lower-gravity ($\chi^2_\nu>4.90$) and/or later types ($\chi^2_\nu>4$).

\subsubsection{Mass and luminosity}
\label{sec:evol_models}
The absolute $H$-band magnitude of HR 1645 B was calculated from the contrast reported in Section~\ref{sec:data} as $m_H = 9.64\pm0.05$\,mag, assuming a negligible correction between the magnitude of the host star in the 2MASS and MKO photometric systems. The flux of the companion and the age posterior distribution for the host star were used in conjunction with the COND03 evolutionary models \citep{Baraffe:2003bj} to derive a model-dependent mass of $110.3_{-3.3}^{+2.0}$\,M$_{\rm Jup}$ and a luminosity of $\log L/L_{\odot}=-2.98\pm 0.02$ using the procedure described in \citet{Chilcote:2017fv}. These errors do not include systematic uncertainties that may be inherent in the model grid, and we assume the age estimate derived from the position of the star on the color-magnitude diagram is not strongly biased by the spectroscopic component described in Section~\ref{sec:rv}. We derive a similar luminosity of $\log L/L_{\odot} = -3.03\pm0.03$ using a $H$-band bolometric correction of $BC(H)=2.50\pm0.07$~mag estimated from an empirical fit to field-gravity objects \citep{Liu:2010cw}. A slightly lower luminosity of $\log L/L_{\odot}=-3.3\pm0.2$ was found using the empirical luminosity-spectral type relationship for field-gravity objects measured by \citet{Filippazzo:2015dv}.

\subsubsection{Visual orbit}
\begin{figure}
\includegraphics[width=1.0\columnwidth]{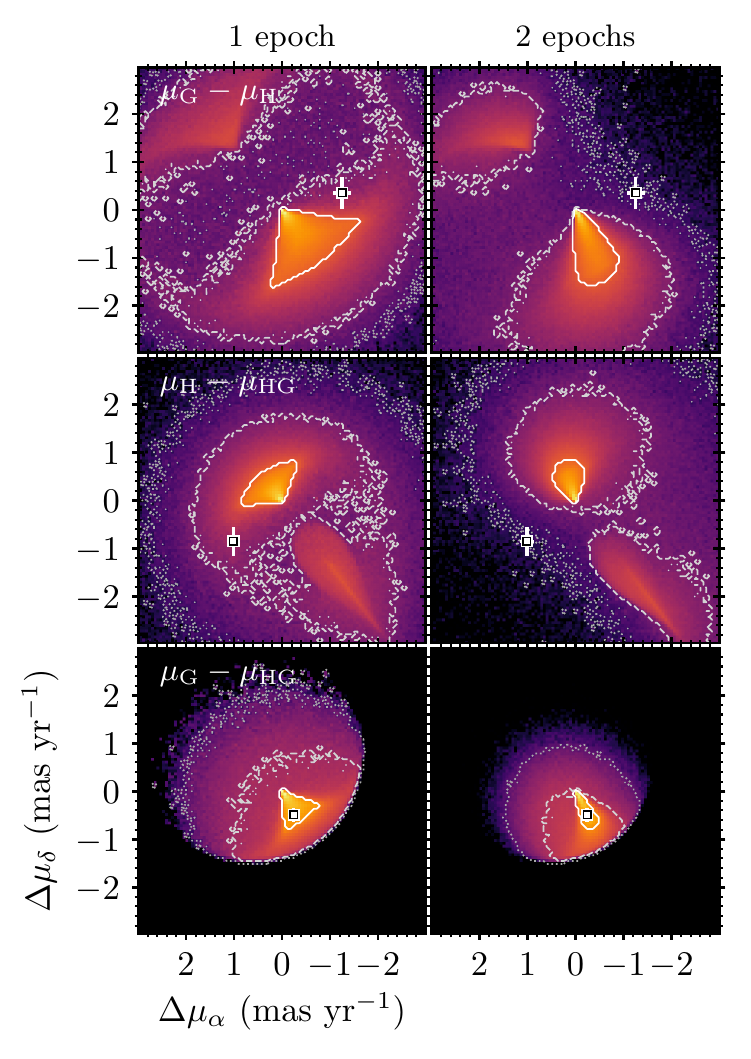}
\caption{Differences between pairs of proper motions of HR 1645 predicted from the fit of the visual orbit of HR 1645 B using the the first (left column) and both (right column) epochs of GPI astrometry. The two-dimensional histograms are plotted on a log scale, with the 1$\sigma$ (solid white), $2\sigma$ (dashed white), and $3\sigma$ (dotted gray contour) credible regions overplotted. The accelerations measured from the {\it Hipparcos} and {\it Gaia} catalogue astrometry (square symbol) appear to be inconsistent with those predicted from the visual orbit fit. \label{fig:accl2d}}
\end{figure}
The visual orbit of HR 1645 B was fit using the relative astrometry from the two GPI datasets given in Section~\ref{sec:data}.  We used the same MCMC sampler described in Section~\ref{sec:accl_fit} to sample the posterior distributions of the semi-major axis $a$, inclination $i$, eccentricity $e$, the sum and difference of the argument of periastron $\omega$ and the longitude of the ascending node $\Omega$, the epoch of periastron $\tau$ (in fractions of the orbital period since the first epoch), the parallax of the system $\pi$, and the mass of the host star $M_1$. The mass of the companion was fixed at 110\,M$_{\rm Jup}$ based on the comparison to evolutionary models in Section~\ref{sec:evol_models}. Standard priors were assumed on the Keplerian elements, and $p(\pi) \propto {\mathcal N}(16.869, 0.092^2)$\,mas based on the {\it Gaia} DR2 measurement, and $p(M_1) \propto {\mathcal N}(1.9,0.1^2)$\,M$_{\odot}$ based on a comparison of the SED of the star to stellar evolutionary models.

The GPI astrometry suggests a counter-clockwise orbit for the companion ($i=71^{+24}_{-25}$\,deg), although inclinations of $i>90^{\circ}$ cannot be excluded at the 1-$\sigma$ level. This is in tension with the inclination estimated from the fit to the {\it Hipparcos} and {\it Gaia} astrometry ($i>90^{\circ}$ for a companion with $P>30$\,yr, see Section~\ref{sec:accl}). To explore this discrepancy further we computed the astrometric signal induced by the companion for each step in the MCMC chain and compared it to the measured values reported in Section~\ref{sec:accl}. We assumed that the instantaneous proper motion of the photocenter at the reference epoch for both missions was a good proxy for the proper motion that would be measured. The wide separation of the companion in late 2018 suggested that there would be minimal acceleration of the proper motion over the {\it Gaia} epoch. This assumption may not be valid for highly eccentric orbits that get significantly closer to the star during the {\it Hipparcos} epoch. The change in the proper motions predicted from the visual orbit fits are shown in Figure~\ref{fig:accl2d}, with the acceleration vectors preferentially aligned with the position angle of the companion in the GPI images. The measured proper motion differences were inconsistent with these predictions at the 3-$\sigma$ level for both $\mu_{\rm G}-\mu_{\rm H}$ and $\mu_{\rm H}-\mu_{\rm HG}$, only $\mu_{\rm G}-\mu_{\rm HG}$ is consistent. This discrepancy is strong evidence that the measured signal can not be entirely ascribed to the the companion resolved in the GPI images. Instead, the signal is either contaminated by an additional companion in the system, or there is a systematic offset between the {\it Hipparcos} or {\it Gaia} astrometry.

\section{HR 1645 A\lowercase{a}A\lowercase{b}}
\label{sec:rv}
While the early-type stars with significant low-amplitude accelerations described in Section~\ref{sec:accl} were screened for known spectroscopic binaries, the suspected multiplicity of HR 1645 was initially missed due to the relatively sparse information regarding the properties of the spectroscopic companion. The Bright Star Catalogue lists HR 1645 as a spectroscopic binary \citep{Hoffleit:1991vr}. The source of this categorization was not given in this catalogue, but was later found within a large bibliography of radial velocities compiled by \citet{Abt:1972tu}. Radial velocity variations were first discovered by \citet{Neubauer:1930jw} (using the alias ``10 G Leporis'' from \citet{Boss:1910uv} that is unfortunately not cross-linked on SIMBAD), and the star was included in a table of spectroscopic binaries discovered during their program \citep{Neubauer:1930ba}. No further information on the properties of the spectroscopic binary was found within the literature, which suggests that the radial velocity variations discovered by \citet{Neubauer:1930jw} were not followed up to measure the spectroscopic orbit.

\subsection{Radial velocities}
\begin{deluxetable}{cccc}
\tablecaption{Radial velocity measurements of HR 1645\label{tbl:rvs}}
\tablehead{\colhead{UT Date} & \colhead{MJD} & \colhead{$v_r$ (km\,s$^{-1}$)} & \colhead{Reference}}
\startdata
1924 Jan 29 & 23813.14 & $-11.1\pm2.88$ & 1\\
1924 Dec 2 & 24121.27 & $30.0\pm2.88$ & 1\\
1926 Feb 5 & 24551.08 & $10.25\pm2.88$\tablenotemark{a} & 1\\
1926 Feb 14 & 24560.05 & $-65.2\pm2.88$\tablenotemark{a} & 1\\
1926 Feb 20 & 24566.06 & $41.6\pm2.88$\tablenotemark{a} & 1\\
1926 Feb 24 & 24570.12 & $40.25\pm2.88$\tablenotemark{a} & 1\\
2002 Oct 5 & 52552.4114 & $27.0\pm2.0$ & 2\\
2009 Feb 9 & 54871.0634 & $25.62\pm5.43$ & 3\\
\enddata
\tablerefs{(1) \citealt{Neubauer:1930jw}; (2) this work; (3) \citealt{Worley:2012ka}.}
\tablenotetext{a}{Weighted average of two measures of the same plate}
\end{deluxetable}

The radial velocities from \citet{Neubauer:1930jw} are given in Table~\ref{tbl:rvs}. The only other radial velocity measurement of the star found in the literature was one derived from VLT/FEROS observations taken in 2009 \citep{Worley:2012ka}, although the quality of the radial velocity measurement is listed as being ``very bad'' in their catalog. We searched the public archives for additional high spectral resolution observations to augment the rather sparse radial velocity record. Only one dataset was found, a 2002 VLT/UVES measurement taken as part of a program to obtain high signal-to-noise ratio high spectral resolution ($R\sim80000$) echelle spectra of stars across the HR diagram \citep{Bagnulo:2003vm}.

The fully-reduced and flux-calibrated UVES spectra were obtained from the UVES Paranal Observatory Projects website\footnote{\url{ https://www.eso.org/sci/observing/tools/uvespop.html}}. We limited our analysis to the six spectra obtained with the blue arm of UVES (``DIC2 437B'') that span 375--495\,nm. This part of the spectrum contains several deep hydrogen lines between 375--440\,nm as well as many shallower lines throughout. We used a grid of synthetic stellar spectra \citep{Allard:2012fp} to identify the temperature, surface gravity, and metallicity that best match the measured spectrum. The synthetic spectra were rotationally broadened assuming a $v\sin i$ of 144 km\,s$^{-1}$ \citep{Zorec:2012il}, and then both the synthetic spectra and the UVES spectrum of HR 1645 were degraded to a resolution of $R\sim 4500$ ($\sim1$\AA\,px$^{-1}$) by convolution with a Gaussian. This grid was linearly interpolated in the three parameters ($T_{\rm eff}$, $\log g$, [Fe/H]) to find the best fit to the observed UVES spectrum through $\chi^2$ minimization. With the best fit parameters in hand, we construct a high resolution ($R\sim80000$, $\sim0.055$\AA\,px$^{-1}$) template from the synthetic spectra.

The radial velocity of HR 1645 was estimated by determining the velocity shift of the high-resolution template spectrum that minimized $\chi^2$ when compared to the UVES spectrum. This process was repeated for each of the six datasets, and once using the entire 375--495\,nm range and once using a more restricted range of 440--475\,nm, avoiding the deep hydrogen lines. We measured a radial velocity of $26.2\pm0.2$\,km\,s$^{-1}$ using the full range and $28.1\pm0.2$\,km\,s$^{-1}$ using the restricted range after applying a barycentric correction. We conservatively adopt $27\pm2$\,km\,s$^{-1}$ as the radial velocity of the star at this epoch. Spectral lines from the companion were not identified in any of the orders, consistent with either a large flux ratio or a negligible velocity differential between the two stars at this epoch.

\subsection{Spectroscopic orbit}
\begin{figure}
\includegraphics[width=1.0\columnwidth]{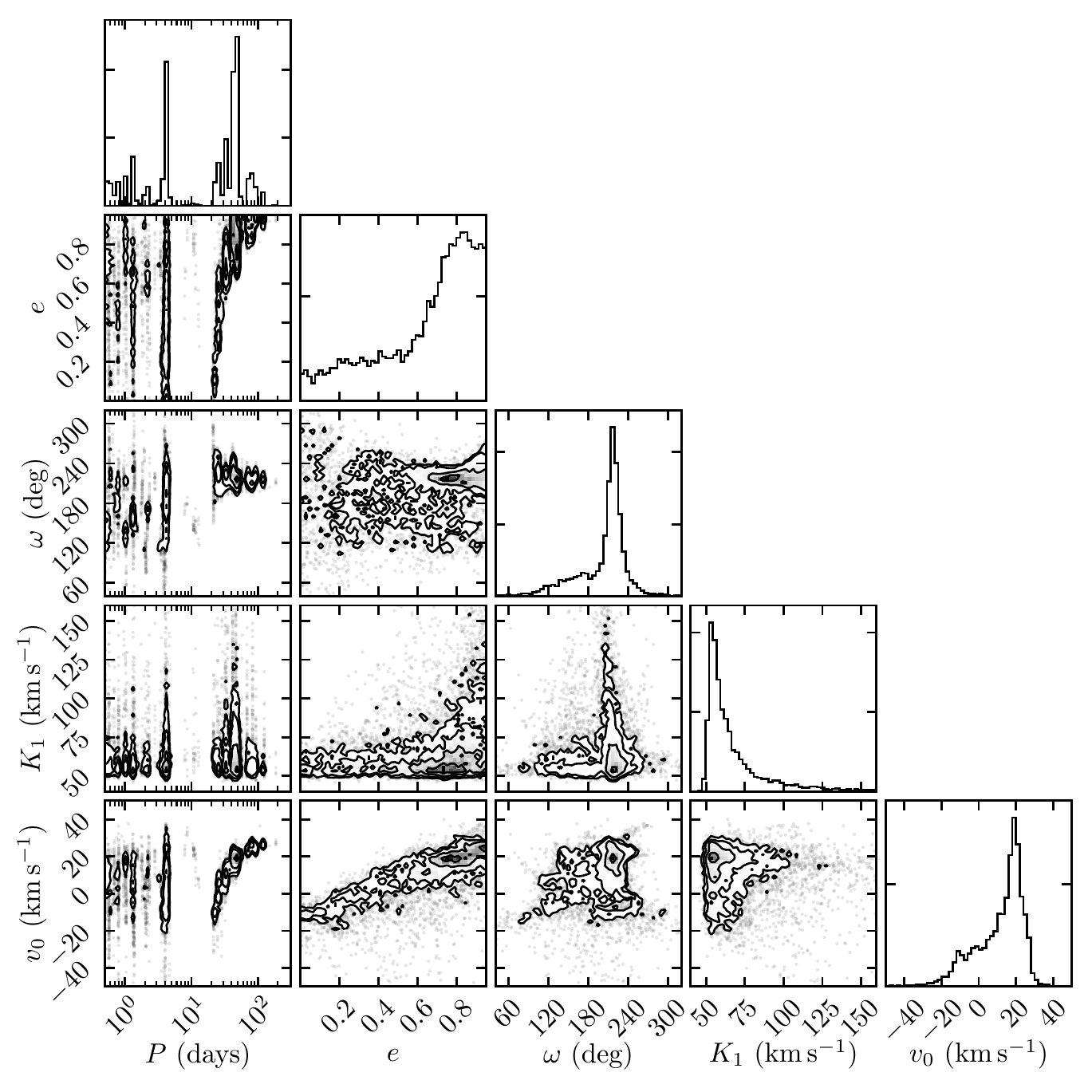}
\caption{Posterior distributions and associated correlations derived from our rejection sampling analysis for four of the Keplerian elements describing the spectroscopic orbit ($P$, $e$, $\omega$, $K_1$) and the radial velocity of the HR 1645 barycenter ($v_0$). The period posterior distribution is highly multimodal due to the sparse sampling of the orbit. \label{fig:rv_corner}}
\end{figure}
\begin{figure}
\includegraphics[width=1.0\columnwidth]{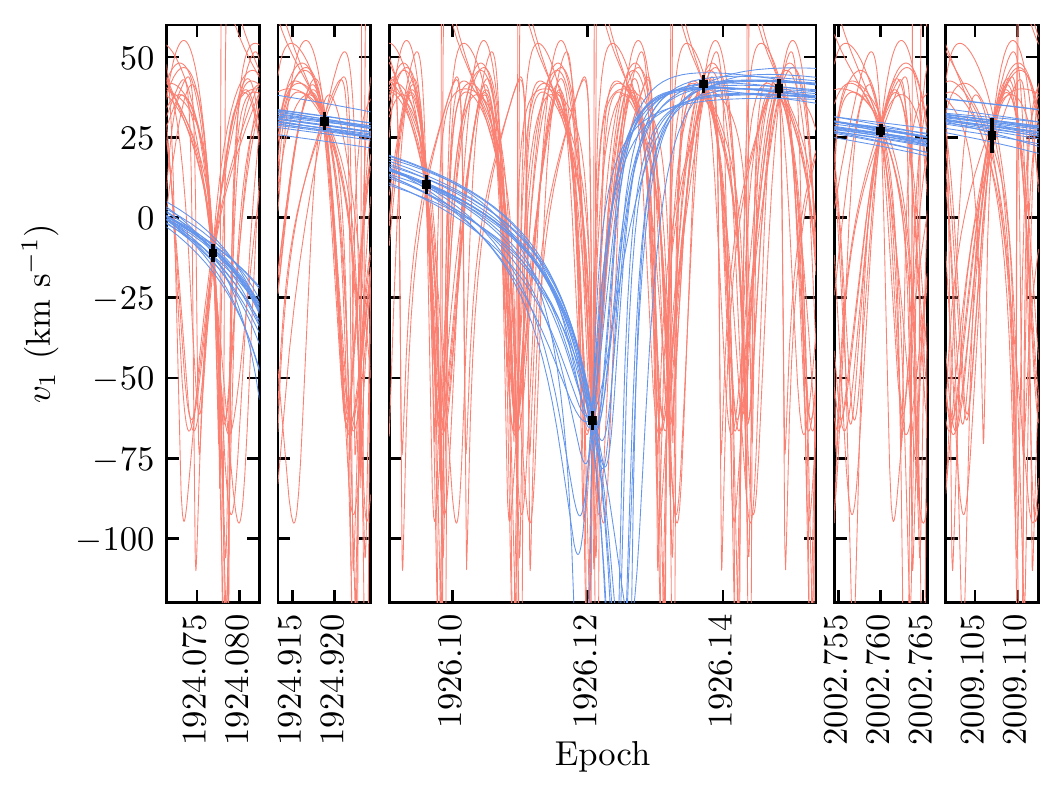}
\caption{Twenty random orbits drawn from our rejection sampling analysis with periods of $3.9\pm0.5$\,d (red) and $46.2\pm0.5$\,d (blue), and the radial velocity measurements from Table~\ref{tbl:rvs} (black squares). Dates without measurements have been omitted due to the 85 years between the first and last epoch. The central panel is 25 days across and contains four epochs, the others are four days across and each contain one epoch. \label{fig:rv_orbit}}
\end{figure}
We used rejection sampling \citep{PriceWhelan:2017br} to efficiently sample the posterior distributions of the orbital elements of the spectroscopic binary. $2^{30}$ ($\sim10^9$) samples were drawn from the prior distributions of the period $P$, eccentricity $e$, argument of periastron $\omega$, and epoch of periastron $\tau$. Prior distributions were uniform in $\log P$ (between 0.5 and 1000 days), $e$ (between 0 and 0.95), $\omega$, and $\tau$. In this framework the radial velocity semi-amplitude $K_1$ and the system velocity $v_0$ are computed analytically for each sample.

Only 7051 of the $2^{30}$ samples were consistent with the radial velocity measurements in Table~\ref{tbl:rvs}. The resulting posterior distributions are shown in Figure~\ref{fig:rv_corner}. The period distribution is multimodal with two pronounced peaks at 3.9\,d and 46.2\,d. Spectroscopic orbits drawn from the posterior distribution at these two periods are plotted in Figure~\ref{fig:rv_orbit}, demonstrating two of the most likely families of orbits. The short-period (4\,d) orbits are uniformly distributed in eccentricity, whereas those with longer periods (46\,d) preferentially have higher eccentricities ($e\sim0.8$). The radial velocity of the HR 1645 barycenter is correlated with the eccentricity of the orbit, and is poorly constrained when considering all allowed orbits ($v_0 = 15.1_{-18.7}^{+8.0}$\,km\,s$^{-1}$). The velocity is similarly poorly constrained for orbits with $\sim$4\,d periods ($v_0 = 1.8_{-13.0}^{+15.6}$\,km\,s$^{-1}$), but is better constrained for those with $\sim$46\,d periods ($v_0 = 18.8_{-2.6}^{+2.2}$\,km\,s$^{-1}$).

We re-evaluated the membership of kinematic associations using these new systemic velocities and the BANYAN web tool \citep{Gagne:2018jj} and found a non-negligible chance of membership of the 149\,Myr \citep{Bell:2015gw} AB Doradus moving group. The membership probability was strongly dependent on the velocity used; ranging from 50--67\,\%, 33--45\,\%, and 0.3--1.8\,\% using the three velocities described previously. The star is most likely not a member of the AB Dor moving group, with  probabilities $\gtrsim90$\,\% typically used as a threshold to assign membership (e.g., \citealp{Gagne:2018jj}). Further spectroscopic monitoring of this system to precisely measure the systemic velocity will be necessary before membership of any moving group can be ascribed based on kinematics alone.

\subsection{Companion mass limits}
While the mass of the close companion cannot be directly measured for a single-lined spectroscopic binary, limits on the mass can be estimated from the spectroscopic orbit. For each orbit found via rejection sampling we computed the mass function $f\left(m\right)$ as
\begin{equation}
\begin{split}
    f\left(m\right) = \frac{K_1^3 P}{2\pi G}\left(1-e^2\right)^{3/2}&=\frac{M_2^3}{\left(M_1+M_2\right)^2}\sin^3i\\
    &=\frac{q^3}{\left(1+q\right)^2}M_1\sin^3i.\\
    \end{split}
\end{equation}
We then computed the minimum values of $q$, and thus $M_2$, by fixing $M_1$ to 1.9\,M$_{\odot}$ and assuming an inclination of 90$^{\circ}$. 98\% of orbits had a minimum mass of $M_2 > 0.3$\,M$_{\odot}$ and 10\% had $M_2>1.9$, typically those with a long period and low eccentricity. The maximum mass of the companion is harder to estimate. The lack of spectral lines from the contemporary spectroscopic datasets is not informative; the derived radial velocities for these epochs are close to the systemic velocity, so the velocity differential between the two components would have been small. \citet{Neubauer:1930jw} do not comment on the presence of additional lines in the spectrum in any of the plates that they analyzed. The difference between the velocity of the two stars in the 1926 February 14 plate should have been $\gtrsim$160\,km\,s$^{-1}$. If the stars were of a similar spectral type the H$\gamma$ absorption line of the two stars would have been separated by 2.3\,\AA, significantly greater than the stated precision of their measurements of 0.04\,\AA. Indeed, \citet{Neubauer:1930jw} report the detection under poor conditions of the spectral lines of both components of the $\mu$ Chamaeleontis system with a velocity differential of 170\,km\,s$^{-1}$. If we conservatively assume that the spectral lines of a companion with with a $V$-band flux ratio of three would have been detected, we can place an upper limit on the mass of a stellar companion at $\sim$1.4\,M$_{\odot}$. There is also the possibility that the companion is a white dwarf, with a similar upper limit of $1.44$\,M$_{\odot}$. 

\subsection{Effect on astrometric measurements}
\begin{figure}
\includegraphics[width=1.0\columnwidth]{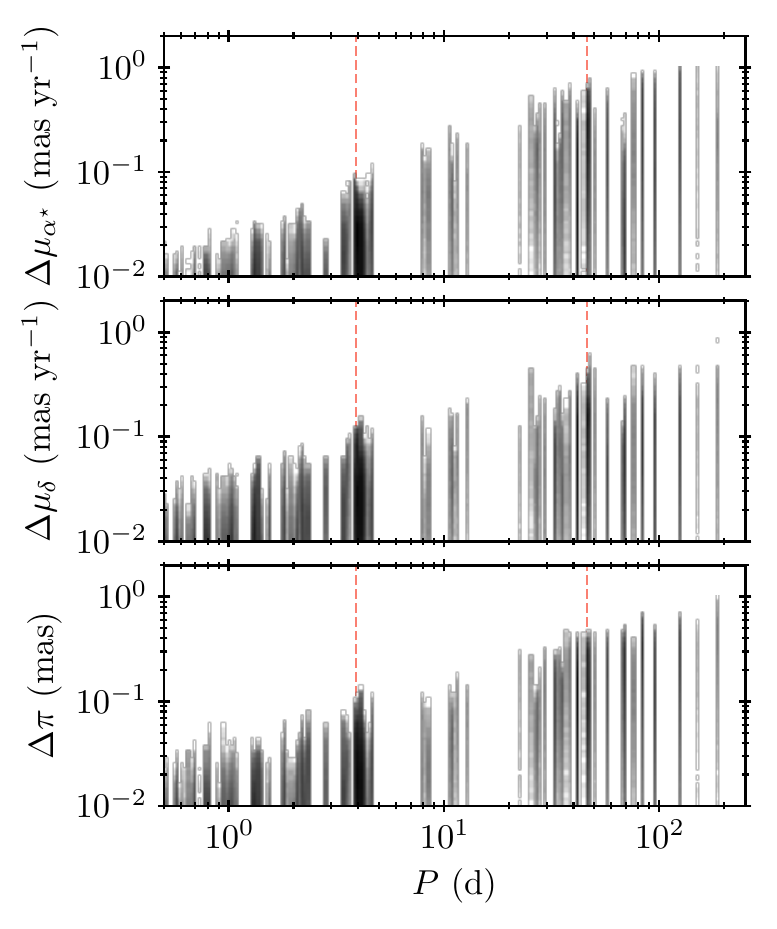}
\caption{Predicted bias in the measurement of $\mu_{\alpha^\star}$ (top), $\mu_{\delta}$ (middle), and $\pi$ (bottom) as a function of orbital period for the spectroscopic binary. The two most probable periods for the binary are highlighted (red dashed lines). The histograms bins are logarithmically scaled to highlight a wide range of orbital periods. The magnitude of the bias at longer orbital periods is comparable to the $\sim1$\,mas\,yr$^{-1}$ astrometric acceleration measured between the {\it Hipparcos} and {\it Gaia} missions. \label{fig:accl_bias}}
\end{figure}
A massive companion with a short orbital period and a large flux ratio can cause motion of the photocenter that can either be detected directly by {\it Hipparcos} and {\it Gaia} (e.g., \citealp{Pourbaix:2000et}), or lead to a spurious measurement of the proper motion of the barycenter of the system due to the aliasing of the observations. In both catalogues the star was fit using a five-parameter astrometric model. The goodness of fit statistic reported in the {\it Hipparcos} catalogue of $F2=2.84$ was worse than median ($F2=1.5_{-1.5}^{+4.5}$) for stars of a similar magnitude, but within the 1$\sigma$ range. The relatively poor goodness of fit may be due to the photocenter motion during the {\it Hipparcos} measurements. The goodness of fit of the {\it Gaia} measurement is much worse at $F2=55.6$, but this is within the 1$\sigma$ range for stars of a similar magnitude ($F2=37.6_{-19.9}^{+31.8}$). The significant difference between the goodness of fit from the two catalogues is likely due to errors induced by the saturation of bright stars on the {\it Gaia} detectors. 

We quantified the potential bias on the proper motion measurement caused by the spectroscopic binary by simulating {\it Hipparcos} measurements of the motion of the photocenter of the system. As the orbit has not been well determined, we used a Monte Carlo algorithm to determine the plausible range of amplitudes of this bias. For each of the 7051 orbits found via rejection sampling we generated $10^3$ astrometric orbits distributed uniformly in $\cos i$, $\Omega$, and $\tau$ (the phasing of the orbit having been lost since the mid-1920s). Approximately half of the generated orbits had $M_2>1.44$\,M$_{\odot}$ and were discarded.

For each of these generated orbits we predicted the motion of the photocenter by combining the proper motion, parallactic motion, and the reflex motion induced by the orbiting companion. The proper motion and parallax of the system barycenter were fixed at ($25$, $-38$)\,mas\,yr$^{-1}$ and $17$\,mas, respectively. The semi-major axis of the orbit of the photocenter around the barycenter was computed as $a_p = (B-\beta)a$, as above. Measurements were simulated at the epochs of the {\it Hipparcos} observations of HR 1645 \citep{vanLeeuwen:2007du}. We then fit a simple five-parameter astrometric model to this simulated dataset using an amoeba-simplex optimization algorithm \citep{Nelder:1965in}. Measurements in the $\alpha^{\star}$ and $\delta$ directions were weighted according to the direction of the scan angle for each epoch. The difference between the proper motion and parallax of the system barycenter and those recovered from the five-parameter fit are plotted as a function of orbital period in Figure~\ref{fig:accl_bias}. The maximum bias in each of the parameters is linearly proportional to the orbital period of the companion. At 4\,d the bias is not significant relative to the catalogue uncertainties. As the period of the binary increases, so does the semi-major axis of the photocenter and thus the magnitude of the astrometric signal induced by the binary. We find a maximum bias on the proper motion measurements of $\sim$0.5\,mas\,yr$^{-1}$ for orbits with a 46\,d period, well above the formal uncertainties on these parameters, and similar in magnitude to the astrometric acceleration of the star measured between the {\it Hipparcos} and {\it Gaia} catalogues.

\section{Conclusion}
We have conducted a detailed study of the HR 1645 system, the primary of which has a significant but low-amplitude acceleration measured in its proper motion between the {\it Hipparcos} and {\it Gaia} missions. We used high-contrast imaging observations to discover a wide-orbit M8 stellar companion, and literature radial velocities to place the first constraints on the spectroscopic orbit of the massive short-period companion. The preliminary fit of the visual orbit of the wide-orbit M8 stellar companion suggests it is unlikely to be inducing the astrometric acceleration of the host star. Instead, it is possible that the aliasing of the photocenter orbit of the short period companion is responsible for the difference in the proper motion between the {\it Hipparcos} and {\it Gaia} catalogues. The nature of the inner companion cannot be determined from the available data. We can place only limited constraints on the mass, and we can infer that the magnitude difference must be non-negligible due to the detected astrometric signal. Future spectroscopic observations of this system could rapidly constrain the orbit of the inner companion, potentially allowing for the astrometric signal induced by this companion over the {\it Hipparcos} and {\it Gaia} missions to be subtracted, leaving only the astrometric acceleration caused by wide companion.

Future targeted searches for wide-orbit companions using a combination of these two catalogues have the potential to reveal a significant population of substellar companions that are amenable to spectroscopic characterization. In this study we have demonstrated how the presence of an additional companion in the system can lead to a spurious detection of an astrometric acceleration that is consistent with a low-mass companion at a wide separation. We were fortunate in this case that the direction of the astrometric acceleration was roughly orthogonal to the position angle of the resolved companion, and that a limited radial velocity record existed in the literature. The contaminating signal does not necessarily have to be from a stellar companion. For example, dynamical masses of wide-orbit planetary-mass companions inferred from an astrometric acceleration and a poorly constrained visual orbit may be biased by the presence of additional substellar companions interior to current sensitivity limits. 

\acknowledgments
Supported by NSF grants AST-1411868 (R.D.R., E.L.N., K.B.F., B.M., and J.P.), AST-141378 (G.D.), AST-1518332 (R.D.R., J.J.W., T.M.E., J.R.G., P.G.K.), and AST1411868 (J.H., J.P.). Supported by NASA grants NNX14AJ80G (R.D.R., E.L.N., S.C.B., B.M., F.M., and M.P.), NSSC17K0535 (R.D.R., E.L.N., B.M., J.B.R.), NNX15AC89G and NNX15AD95G (R.D.R., B.M., J.E.W., T.M.E., G.D., J.R.G., P.G.K.). This work benefited from NASA's Nexus for Exoplanet System Science (NExSS) research coordination network sponsored by NASA's Science Mission Directorate. J.R is supported by the French National Research Agency in the framework of the Investissements d’Avenir program (ANR-15-IDEX-02), through the funding of the ``Origin of Life'' project of the University Grenoble-Alpes. Portions of this work were performed under the auspices of the U.S. Department of Energy by Lawrence Livermore National Laboratory under Contract DE-AC52-07NA27344. Based on observations obtained at the Gemini Observatory, which is operated by the Association of Universities for Research in Astronomy, Inc., under a cooperative agreement with the NSF on behalf of the Gemini partnership: the National Science Foundation (United States), National Research Council (Canada), CONICYT (Chile), Ministerio de Ciencia, Tecnolog\'{i}a e Innovaci\'{o}n Productiva (Argentina), Minist\'{e}rio da Ci\^{e}ncia, Tecnologia e Inova\c{c}\~{a}o (Brazil), and Korea Astronomy and Space Science Institute (Republic of Korea). This work has made use of data from the European Space Agency (ESA) mission {\it Gaia} (\url{https://www.cosmos.esa.int/gaia}), processed by the {\it Gaia} Data Processing and Analysis Consortium (DPAC, \url{https://www.cosmos.esa.int/web/gaia/dpac/consortium}). Funding for the DPAC has been provided by national institutions, in particular the institutions participating in the {\it Gaia} Multilateral Agreement. This research has made use of the SIMBAD database and the VizieR catalog access tool, both operated at the CDS, Strasbourg, France. This research has made use of the ``Modern Mean Dwarf Stellar Color and Effective Temperature Sequence'' available at \url{http://www.pas.rochester.edu/~emamajek/EEM_dwarf_UBVIJHK_colors_Teff.txt}. This research has benefited from the SpeX Prism Library (and/or SpeX Prism Library Analysis Toolkit), maintained by Adam Burgasser at \url{http://www.browndwarfs.org/spexprism}, the IRTF Spectral Library, maintained by Michael Cushing, the Brown Dwarfs in New York City database led by Jackie Faherty, Emily Rice, and Kelle Cruz, and the Montreal Brown Dwarf and Exoplanet Spectral Library, maintained by Jonathan Gagn\'{e}.

\facility{Gemini:South (GPI)}

\software{Astropy \citep{TheAstropyCollaboration:2013cd},  
          Matplotlib \citep{Hunter:2007ih}}

\bibliographystyle{aasjournal}   
\bibliography{refs}

\end{document}